\documentclass[%
preprint,
aps,
prb,
showpacs,
showkeys,
superscriptaddress,
amsmath,amssymb,
longbibliography,
]{revtex4-2}

\usepackage[utf8]{inputenc}
\usepackage{multirow}
\usepackage{graphicx}
\usepackage{bm}
\usepackage{mathtools}
\usepackage{dcolumn}
\usepackage{picture}
\usepackage{physics}
\usepackage[colorlinks=true,urlcolor=blue,linkcolor=blue,citecolor=blue]{hyperref}
\usepackage{etoolbox}
\usepackage{float}
\usepackage{natbib}
\usepackage{times}
\usepackage{color}
\usepackage{xcolor}
\usepackage{soul}
\usepackage[per-mode=symbol]{siunitx}
\usepackage{amssymb}
\usepackage{amsmath,latexsym}
\usepackage[normalem]{ulem}
\usepackage{hyperref}
\usepackage{url}

\begin{document}

	\title{Crystal growth, magnetic, and transport properties of the candidate altermagnet Cr$_{1/4}$NbS$_{2}$}
	
	\author{Nashra Pistawala}
%    \email[email:]{pistawalanashra(at)gmail.com}
	\affiliation{Physics department, University of Warwick, Coventry, CV47AL, United Kingdom}
	
         \author{Martin R. Lees}
	\affiliation{Physics department, University of Warwick, Coventry, CV47AL, United Kingdom}
   
	\author{Geetha Balakrishnan}
	\affiliation{Physics department, University of Warwick, Coventry, CV47AL, United Kingdom}

    \author{Daniel A. Mayoh}
    \email[email:]{D.Mayoh.1@warwick.ac.uk}
	\affiliation{Physics department, University of Warwick, Coventry, CV47AL, United Kingdom}

	\date{\today}
	
	\begin{abstract}

Altermagnetism is a rapidly emerging paradigm in condensed matter magnetism, yet the number of experimentally confirmed altermagnetic candidates remains remarkably limited. Cr$_{1/4}$NbS$_2$ has long been theoretically proposed as a promising altermagnet, but there are no reports on experimental investigation of single crystals. Here, we report the successful growth of high-quality single crystals of Cr$_{1/4}$NbS$_2$ and present its structural, thermal, magnetic, and transport properties. Single-crystal X-ray diffraction provides detailed crystallographic information and confirms the hexagonal crystal symmetry ($P6_3/mmc$), one of the criteria for altermagnetic behavior. Resistivity and heat capacity measurements unambiguously identify an antiferromagnetic N\'{e}el temperature of $T_N \sim 141$\,K. Magnetotransport measurements reveal non-trivial features in the magnetoresistance and Hall response, indicative of quantum diffusive transport phenomena. Cr$_{1/4}$NbS$_2$, synthesized here for the first time in single-crystalline form, therefore serves as an experimental platform for probing the rich spectrum of phenomena predicted for altermagnetic materials.

\end{abstract}

\maketitle
\label{Introduction}
 
\section{Introduction}
\label{Intro}
 Magnetism has long been classified under two canonical paradigms: ferromagnetism (FM), characterized by net spontaneous magnetization, and antiferromagnetism (AFM), defined by compensated magnetic sublattices and zero net magnetization~\cite{doi:10.1126/science.174.4013.985}. However, a third class of magnetic order, termed \textit{altermagnetism} (AM), has recently emerged as a transformative concept that challenges this binary classification~\cite{PhysRevX.12.031042, PhysRevX.12.040002}. Altermagnets are collinear, spin-compensated magnets that nevertheless exhibit momentum-dependent spin splitting in their electronic band structure, a property previously thought exclusive to ferromagnets and non-collinear antiferromagnets. This spin splitting arises not from spin-orbit coupling but from the real-space crystal symmetry connecting opposite-spin sublattices via rotational operations (proper or improper and symmorphic or non-symmorphic), rather than the inversion or translation symmetries characteristic of conventional antiferromagnets~\cite{PhysRevX.12.040501, PhysRevX.12.011028, https://doi.org/10.1002/adma.202211966, Krempasky2024, PhysRevLett.132.176702, PhysRevLett.133.236703, Yan2024}. As a consequence, altermagnets combine the high stability and zero stray fields of antiferromagnets with the spin-polarized transport properties of ferromagnets, positioning them as promising candidates for next-generation spintronics~\cite{Jungwirth2025}.
 
Although the experimental studies of altermagnetism is still in its early stages, recent advances in experimental observations and theoretical calculations have revealed several exotic physical phenomena, including nonrelativistic lifted Kramers degeneracy ~\cite{Krempasky2024, https://doi.org/10.1002/adma.202314076}, anomalous spin transport ~\cite{PhysRevLett.130.036702, doi:10.1126/sciadv.adn0479}, magneto-optical effects~\cite{PhysRevB.106.064435, PhysRevLett.132.176701}, and chiral magnons~\cite{PhysRevLett.131.256703, PhysRevLett.133.156702}. The theoretical framework for altermagnetism has rapidly matured, with first-principles calculations predicting huge number of candidate materials spanning insulators, semiconductors, and metals~\cite{https://doi.org/10.1002/adfm.202409327, Chen2025, PhysRevMaterials.9.064403, GUO2023100991, 10.1093/nsr/nwaf066, Song2025, doi:10.1021/acsorginorgau.4c00064}. Among the most widely studied candidates are RuO$_2$~\cite{https://doi.org/10.1002/advs.202400967}, $\alpha$-MnTe~\cite{PhysRevLett.134.086701}, CrSb~\cite{Yang2025}, MnF$_2$~\cite{Lovesey2026}, Mn$_5$Si$_3$~\cite{Badura2025} and La$_2$CuO$_4$~\cite{Saunderson2025} which collectively span a broad range of N\'{e}el temperatures ($T_{\mathrm{N}} = 69$--$400$~K) and spin-splitting magnitudes ($10$--$1400$~meV)~\cite{PhysRevX.12.040501}. RuO$_2$ has attracted the most intense scrutiny as the archetypal altermagnet, due to its rutile crystal structure and predictions of giant spin splitting~\cite{PhysRevLett.132.056701}. However, the magnetic ground state of RuO$_2$ remains vigorously debated: whereas neutron diffraction and transport studies report signatures of altermagnetism~\cite{PhysRevLett.118.077201, PhysRevLett.122.017202}, muon spin rotation ($\mu$SR) measurements find no evidence of broken time-reversal symmetry, leaving the question unresolved if RuO$_2$ even possesses any magnetic order ~\cite{Kessler2024}. $\alpha$-MnTe has been shown to exhibit anisotropic band splitting of ~0.8 eV arising from time-reversal symmetry breaking, as observed using angle-resolved photoemission spectroscopy (ARPES) in combination with first-principles calculations, providing the first direct experimental evidence of altermagnetic band splitting~\cite{PhysRevB.109.115102}. CrSb on the other hand shows a significant spin splitting of 0.6 eV detected using soft X-ray ARPES with position of the electronic bands just below the Fermi energy making it an ideal candidate for spintronics devices~\cite{Reimers2024}. Despite these advances, the number of experimentally confirmed altermagnetic candidates remains remarkably small compared to the large pool of theoretically proposed systems. This is perhaps surprising, given that recent symmetry analyses suggest altermagnetism to be the default behavior in antiferromagnets, with spin splitting being ubiquitous rather than exceptional~\cite{Spaldin2026}. The scarcity of confirmed candidates thus reflects not a true rarity in nature, but rather the demanding experimental requirements needed to unambiguously identify altermagnetic signatures, and hence pressing need for new, well-characterized candidate materials.

Cr$_{1/4}$NbS$_2$ is a member of the family of transition metal-intercalated transition metal dichalcogenides (TMDs) and has recently been predicted to be a potential altermagnet~\cite{GUO2023100991, https://doi.org/10.1002/adfm.202409327, Chen2025}. Band structure calculations by Guo \textit{et al.}~\cite{GUO2023100991} reveal significant spin splitting in the electronic bands even in the absence of spin--orbit coupling, a key hallmark of altermagnetic behavior. Notably, Cr$_{1/4}$NbS$_2$ is among the few proposed altermagnetic candidates that exhibit metallic conductivity, making it particularly attractive for magnetotransport studies~\cite{GUO2023100991}. Further motivation comes from a recent experimental study on Co$_{1/4}$NbSe$_2$, another member of the same $TM_{1/4}$Nb$X_2$ ($TM =$~transition metal, $X =$~S, Se, Te) family, in which clear evidence of altermagnetic band splitting was demonstrated through combined DFT calculations and ARPES measurements~\cite{Regmi2025}. Cr$_{1/4}$NbS$_2$ has thus emerged as a theoretically compelling altermagnetic candidate warranting detailed experimental investigation. However, experimental studies on this material remain scarce, and all prior investigations have been conducted exclusively on polycrystalline specimens~\cite{VANLAAR1971154, Marseglia01011983}. The orientation dependent magnetic and magnetotransport studies are essential for probing the anisotropic electronic and magnetic properties predicted for altermagnets. In this work, we report the successful growth of high-quality single crystals of Cr$_{1/4}$NbS$_2$ by the chemical vapor transport technique and present a comprehensive investigation of their structural, chemical, magnetic, and transport properties. Using single-crystal X-ray diffraction we obtain the detailed crystallographic information and further confirm the hexagonal crystal symmetry required to observe the altermagnetic behavior. Electrical resistivity and heat capacity measurements unambiguously establish an antiferromagnetic N\'{e}el temperature of $T_N \sim 141$~K, while magnetotransport measurement shows unusual features indicative of diffusive quantum transport phenomenon. The successful synthesis of Cr$_{1/4}$NbS$_2$ single crystals and the systematic experimental investigation of their transport and magnetic properties presented here pave the way for future exploration of altermagnetic behaviour in this compound.

 \section{Experimental Methods}
	\label{Exp}
Single crystals of Cr$_{1/4}$NbS$_2$ were grown by the chemical vapor transport (CVT) method, the details of which are described in Sec.~\ref{CG}. The quality and orientation of the as-grown single crystals were determined by X-ray Laue diffraction using a Photonic Science Laue camera system operating in back-scattering geometry. Single-crystal X-ray diffraction (SCXRD) measurements were performed using a Rigaku Oxford Diffraction Synergy S diffractometer, employing mirror monochromated Mo-K$_\alpha$ radiation ($\lambda = 0.70173$~\AA) generated by a microfocus sealed tube PhotonJet X-ray source, with diffracted intensities recorded on a HyPix hybrid photon-counting detector. A platelet-shaped single-crystal specimen of dimensions approximately $0.2 \times 0.2 \times 0.2$~mm$^3$, cut from a larger crystal, was measured at 100~K. The intensities were integrated using the CrysAlisPro~\cite{Crystalispro} and the structure solution and refinement were carried out using \textsc{Olex2}~\cite{olex2}. The chemical homogeneity and average elemental composition of the single crystals were confirmed by scanning electron microscopy (SEM) coupled with energy-dispersive X-ray spectroscopy (EDX), carried out using a ZEISS GeminiSEM 500 field-emission SEM (FESEM) equipped with an Oxford Instruments EDX detector. EDX spectra were collected at 15--20 different points across a crystal specimens to obtain an average elemental composition. This was further repeated for 4-5 crystal specimens in order to confirm the homogeneity of the crystals grown in this batch. Heat capacity measurements were performed in a Quantum Design Physical Property Measurement System (PPMS) using the $2\tau$ thermal relaxation method at temperatures from 2 to 270~K in zero applied field. Since the individual single crystals were too small for reliable standalone measurement, several crystals were stacked together using Apiezon N grease to achieve a total sample mass of approximately 12--13~mg. A small error in the absolute heat capacity values may arise due to an uncertainty in the mass of the N grease used to stack the crystals together. DC magnetization measurements were performed as a function of temperature (1.8--300~K) and applied magnetic field (0--7~T) using a Quantum Design Magnetic Property Measurement System (MPMS) magnetometer. Electrical resistivity and Hall measurements were carried out as a function of temperature (1.8--300~K) and magnetic field (0--9~T) using the AC transport option of a Quantum Design PPMS, with excitation currents in the range 1--10~mA at a frequency of 113~Hz for all the transport measurements.

	\section{Results and Discussion}
	\label{RD}
	\subsection{Crystal growth of Cr$_{1/4}$NbS$_2$}
	\label{CG}

\begin{figure}
		\centering
		\includegraphics[width= 1.0\columnwidth]{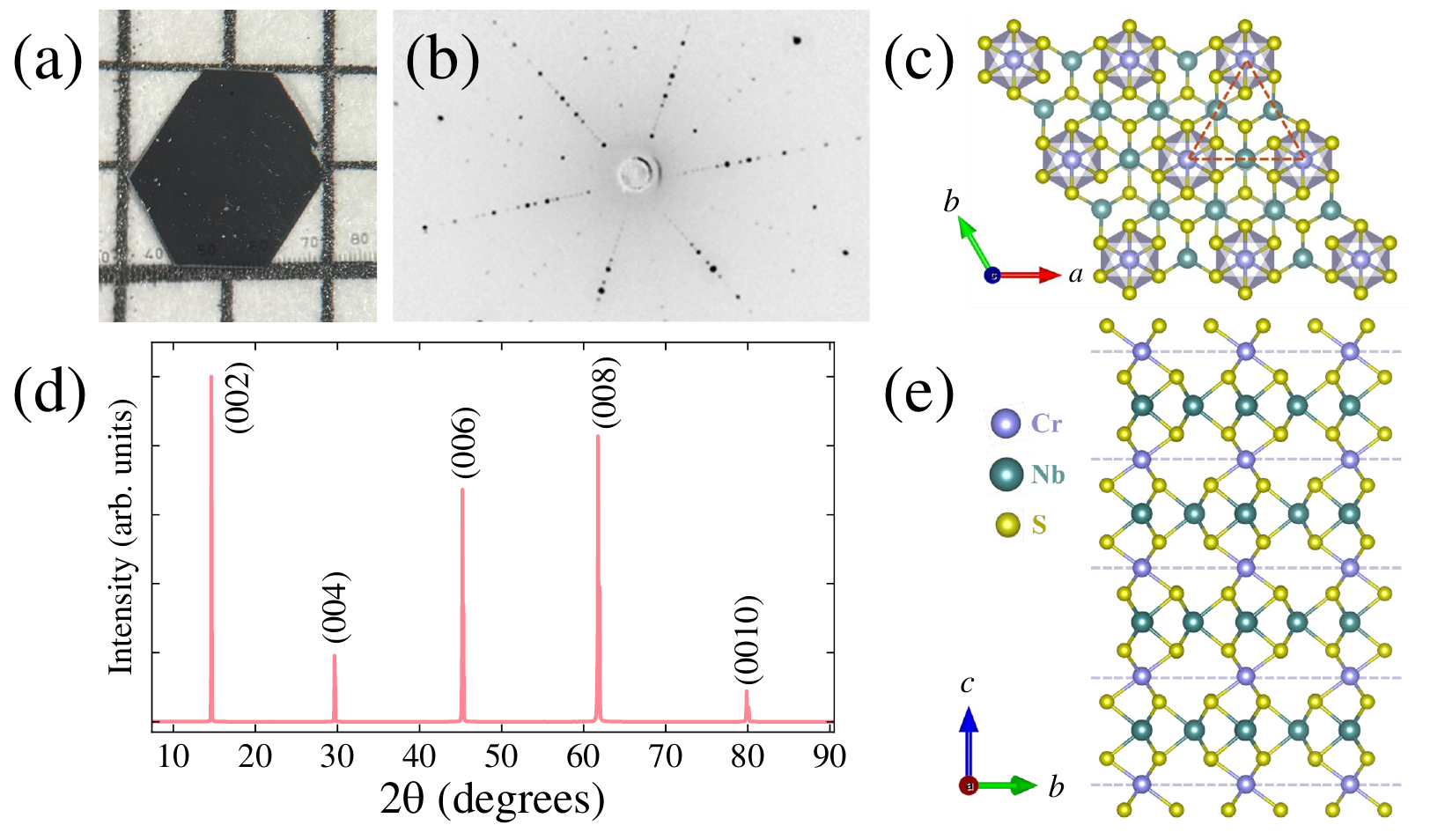}
		\caption {(a) Image of an as-grown single crystal of Cr$_{1/4}$NbS$_2$; each grid square corresponds to 1\,mm~$\times$~1\,mm. (b) Back-scattering Laue diffraction pattern collected from the crystal shown in (a) confirms the crstallographic $ab$-plane lies in the plane of the crystal. (c) X-ray diffraction pattern recorded in Bragg--Brentano geometry, showing dominant $(0 0 l)$ reflections, confirming that the plane of the crystal shown in (a) corresponds to the $ab$-plane. Crystal structure of Cr$_{1/4}$NbS$_2$ visualized using VESTA~\cite{momma2011vesta}: (d) projection in the $ab$-plane showing the triangular arrangement of Cr$^{3+}$ ions (orange dashed lines), and (e) projection along the $c$-axis illustrating the alternating 2H-NbS$_2$ and Cr$^{3+}$ intercalation layers, with Cr$^{3+}$ occupying the $2a$ Wyckoff site (purple dashed lines).}
	\label{CS}
 \end{figure}

Single crystals of Cr$_{1/4}$NbS$_2$ were grown by the chemical vapor transport method using a polycrystalline precursor of Cr$_{1/4}$NbS$_2$. The polycrystalline precursor was prepared via solid-state synthesis. Stoichiometric quantities of chromium (powder, 99\%, Thermo Scientific), niobium (powder, 99.99\%, Thermo Scientific), and sulfur (powder, 99.99\%, Thermo Scientific) were thoroughly ground together inside an argon-filled glovebox and sealed in an evacuated quartz ampoule. To avoid a buildup of pressure from sulfur vapor, the sealed ampoule was initially heated slowly to 450~$^{\circ}$C at a rate of 15~$^{\circ}$C/hr in several steps. The ampoule was subsequently heated to 950~$^{\circ}$C, held at this temperature for 100~hr, and finally cooled slowly to room temperature at a rate of 5~$^{\circ}$C/hr. The resulting ingot was thoroughly ground to obtain a fine powder. For the crystal growth, the powder was sealed in a quartz ampoule together with iodine (I$_2$, ACS reagent, $\geq$99.8\%, Sigma-Aldrich) as the transporting agent. The ampoule was then placed in a two-zone furnace with the source zone maintained at 950~$^{\circ}$C and the sink zone at 800~$^{\circ}$C (temperature gradient $\Delta T \approx 150~^{\circ}$C) for a duration of 20 days. The process is similar to that used for Cr$_{1/3}$NbS$_2$~\cite{PhysRevMaterials.6.024407, PhysRevResearch.4.013134}.  Shiny, platelet-like single crystals of approximate dimensions $2 \times 2 \times 0.5$~mm$^3$ were obtained at the sink end of the ampoule after the transport reaction, as shown in Fig.~\ref{CS}(a). The Laue back-scattering image of a representative crystal is shown in Fig.~\ref{CS}(b). The sharp, well-defined Laue spots, corresponding to $(h k 0)$ reflections and obtained with the c-axis oriented perpendicular to the crystal surface, exhibit the six-fold symmetry characteristic of the hexagonal system, confirming the high crystalline  quality of the grown single crystals. Figure~\ref{CS}(c) shows the X-ray diffraction pattern in the plane of the single crystal collected in a Bragg--Brentano geometry, showing only $(0 0 l)$ reflections, confirming the $ab$-plane orientation of the crystal. The homogeneity and chemical composition of the grown single crystals are presented in the Supplemental Material Sec.~A~\cite{supplementary}. The average composition obtained from EDX analysis is Cr$_{0.261(6)}$NbS$_{1.95(2)}$, in good agreement with the nominal stoichiometry of Cr$_{0.25}$NbS$_{2}$ within the experimental uncertainty of the EDX measurements.

	\subsection{Crystal Structure}
	\label{structure}
    %\label{CS}

    Cr$_{1/4}$NbS$_2$ belongs to the family of transition metal-intercalated transition metal dichalcogenides with the general formula Cr$_x$NbS$_2$, which crystallizes in two distinct symmetries depending on the concentration $x$ of the intercalant~\cite{levy1979intercalated, Marseglia01011983}. For $x = 1/3$, the system adopts the noncentrosymmetric chiral space group $P6_322$, where the Cr atoms occupy the 2c Wyckoff position and order into a $(\sqrt{3} \times \sqrt{3}) $R30$^{\circ}$ superstructure~\cite{Parkin01011980, Parkin01011980_2}. The resulting broken inversion symmetry gives rise to a Dzyaloshinskii--Moriya interaction, driving a chiral helimagnetic ground state and, under applied field, a chiral soliton lattice, both of which have been extensively investigated in the literature~\cite{PhysRevLett.108.107202, PhysRevB.92.220412, PhysRevB.96.184423, CAO2020100080, PhysRevMaterials.6.024407}. For $x = 1/4$, the system instead crystallizes in the centrosymmetric space group $P6_3/mmc$, where the Cr atoms occupy the octahedral 2a Wyckoff sites and form a $2 \times 2$ in-plane superstructure~\cite{VANLAAR1971154, Marseglia01011983}. Here, the inversion symmetry is preserved and Cr$_{1/4}$NbS$_2$ is expected to exhibit conventional magnetic ordering.  Single-crystal X-ray diffraction measurements on a crystal specimen grown in this work confirms the $P6_3/mmc$ symmetry, in agreement with the early work of Van Laar~\textit{et al.}~\cite{VANLAAR1971154}. The data collection strategy, crystallographic information, and refinement parameters are summarized in Table~\ref{SCXRD_CD}. The atomic positions, refined occupancy and corresponding isotropic atomic displacement parameters are listed in Table~\ref{Table_AP}. The refined occupancy of Cr$^{3+}$ at the 2a Wyckoff site is marginally below full occupancy, however, it remains 1 to within the experimental error bars. The crystal structure of Cr$_{1/4}$NbS$_2$, visualized using VESTA, is shown in Figs.~\ref{CS} (d, e). The Cr$^{3+}$ ions form a triangular network in the $ab$-plane, as shown in Fig.~\ref{CS}(d), while Fig.~\ref{CS}(e) illustrates the stacking of 2H--NbS$_2$ layers along the $c$-axis with the Cr ions occupying the octahedral (2a - Wyckoff site) interlayer sites. The centrosymmetric space group of Cr$_{1/4}$NbS$_2$ suppresses the Dzyaloshinskii–Moriya interaction that drives helimagnetic order in non-centrosymmetric Cr$_{1/3}$NbS$_2$, and provides a crystallographic framework which is consistent with the symmetry requirements favorable for altermagnetism.

\begin{table}[!]
\caption{\label{SCXRD_CD}%
Crystallographic data and refinement parameters for SCXRD of Cr$_{1/4}$NbS$_2$ single crystal.}
\centering
\begin{ruledtabular}
\begin{tabular}{cc}
Temperature (K) &  100  \\
Crystal system & Hexagonal  \\
Space group & $P6_3/mmc$ \\
 No.  & 194   \\
$a$ (\AA{}) & 6.62280(10)   \\
$c$ (\AA{}) & 11.9573(3)  \\
Volume (\AA{}$^3$) & 454.200(18)  \\
$Z$ & 2 \\
Density (g/cm$^3$) & 4.96 \\
Wavelength (\AA{}) & 0.71073  \\
%Detector distance (mm) & 40  \\
%Scan type & Full sphere ($\phi$, $\omega$ scans) \\
Rotation per image (deg) & 0.2  \\
Exposure time (sec) & 10 \\
$\theta$ (${}^{\circ}$) & \numrange{3.41}{30.45}\\
Index ranges  & $-9 \leq h \leq 9$ \\
& $-9 \leq k \leq 9$ \\
 & $-17 \leq l \leq 17$ \\
%$(\sin(\Theta)/\lambda)_{max}$ (\AA{}$^{-1}$) & 0.743374 \\
Absorption, $\mu$ (mm$^{-1}$) & 7.78  \\
T$_{\mathrm{min}}$, T$_{\mathrm{max}}$ & 0.001, 0.009  \\
Criterion of observability & $I>2\sigma(I)$ \\
No. of  reflections measured & 12285  \\
Unique reflections (obs/all) & 293/301  \\
$R_{\mathrm{int}}$   &0.0622  \\
No. of parameters &19 \\
$R_{\mathrm{F}}$   (obs) &0.0911 \\
$wR_{\mathrm{F}}$ (all) &0.2617 \\
GoF  &1.360 \\
$\Delta\rho_{\mathrm{min}}$, $\Delta\rho_{\mathrm{max}}$(e \AA$^{-3}$) &-1.423, 5.523  \\
\end{tabular}
\end{ruledtabular}
\end{table}

\begin{table}[tbh!]
\caption{\label{Table_AP}%
Atomic coordinates $x$, $y$, $z$ and the equivalent atomic displacement parameter of single crystal Cr$_{1/4}$NbS$_2$ at 100~K. Unique reflections (obs/all) = 293 / 301, criterion of observability: $I>2\sigma(I)$, $R_{\mathrm{F}} = 0.0911$, No. of refined paramateres = 19, Refinement method used: least-squares on F, Space group: $P6_3/mmc$. The anisotropic atomic displacement parameters are shown in the Supplemental Material Sec.~B Table~S2~\cite{supplementary}.}
\centering
\begin{ruledtabular}
\begin{tabular}{c c c c c c c}
Atoms & Wyck & $x$ & $y$ & $z$ & Occ. & $U_{\mathrm{iso}}^{\mathrm{eq}}\left(\mathrm{\AA}^{2}\right)$ \\[1ex]
\hline\\
Cr & 2a & 1.0000 & 1.0000 & 0.5000 & 0.97(4) & 0.0297(13) \\
Nb1 & 2b & 1.0000 & 1.0000 & 0.2500 & 1 & 0.0281(10) \\
Nb2 & 6h & 0.4942(12) & 0.9884(2) & 0.2500 & 1 & 0.0285(10) \\
S1 & 12k & 0.8328(15) & 1.1671(15) & 0.3811(3) & 1 & 0.0289(12) \\
S2 & 4f & 0.3333 & 0.6667 & 0.1152(3) & 1 & 0.0289(13) \\
\end{tabular}
\end{ruledtabular}
\end{table}

	\subsection{Heat Capacity}
        \label{HC}

\begin{figure}[tbh!]
		\centering
		\includegraphics[width= 0.5\columnwidth]{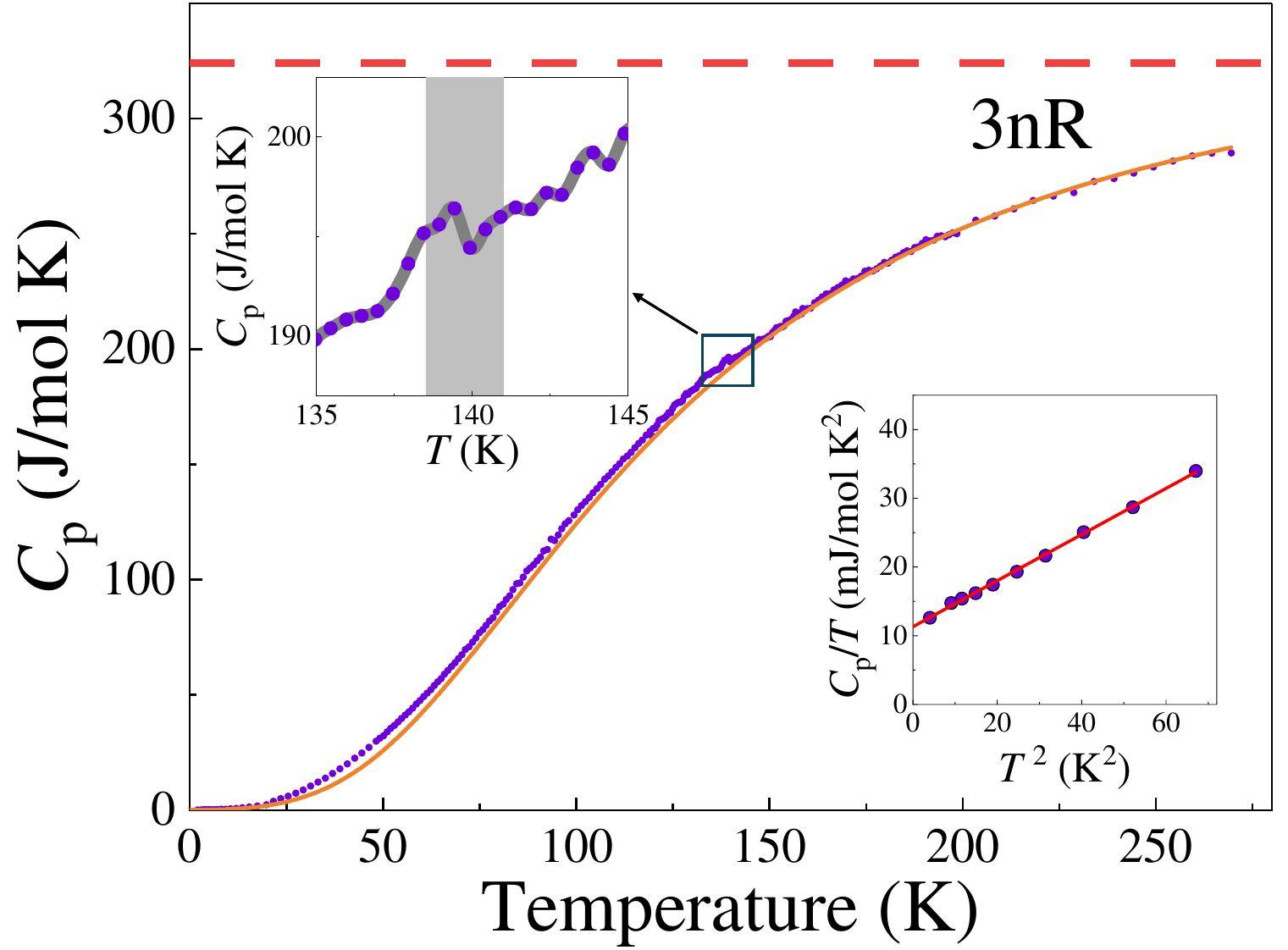}
		\caption {Temperature dependence of the heat capacity of Cr$_{1/4}$NbS$_2$. The solid orange line is a fit to the non-magnetic contribution to the heat capacity, as described in the text. The top-left inset shows a magnified view of the data near the transition temperature $T_{\mathrm{N}}$ of 141(1)~K. The inset bottom right shows the $C_{\mathrm{p}}/T$ vs $T^2$ behavior at low temperature; the red solid is a linear fit to the data (details of which are described in the text).} 
	\label{HCplot}
 \end{figure}

The temperature dependence of the heat capacity for single crystal Cr$_{1/4}$NbS$_2$ is shown in Fig.~\ref{HCplot}. The heat capacity exhibits a small $\lambda$-like anomaly characteristic of a second-order phase transition. A view of the $C_{\mathrm{p}}$ vs\ $T$ data around this anomaly is shown in the top-left inset of Fig.~\ref{HCplot}. This feature signals an antiferromagnetic transition, with the N\'eel temperature, $T_{\mathrm{N}}$, determined to be 141(1)~K, in good agreement with the value obtained from resistivity measurements (see Sec.~\ref{Transport}).  At high temperatures, the heat capacity approaches the classical Dulong--Petit limit, given by $3NR$, where $N$ is the number of atoms per formula unit and $R$ is the universal gas constant. For Cr$_{1/4}$NbS$_2$ ($N=13$), this corresponds to $\sim324$~J~mol$^{-1}$~K$^{-1}$. At low temperatures the data were fit using the relation $C_{\mathrm{p}}(T)/T=\gamma+\beta T^2$,  as shown in the bottom-right inset of Fig.~\ref{HCplot}, where $\gamma$ is the Sommerfeld coefficient (electronic contribution) and $\beta T^3$ is made up of lattice (phonon) and a spin-wave (magnon) contributions. The fit yields $\gamma=11.3(2)$~mJ~mol$^{-1}$~K$^{-2}$ and $\beta=0.336(4)$~mJ~mol$^{-1}$~K$^{-4}$. The value of $\gamma$ is consistent with reports for other members of the family of transition metal intercalated transition metal dichalcogenide~\cite{PhysRevB.108.054418}.
The $C_{\mathrm{p}}(T)$ data were fit between 145 and 270~K using a Debye model including an electronic ($\gamma T$) term, in order to estimate the non-magnetic contribution to the heat capacity, with the fit then extrapolated to 0~K (see Fig.~\ref{HCplot}). The fit gave a Debye temperature $\theta_{\mathrm{D}}$ of 502(3)~K. Subtracting this non-magnetic component gave the magnetic contribution to the heat capacity $C_{\mathrm{mag}}(T)$. See Supplemental Material for more details on estimating $C_{\mathrm{mag}}(T)$~\cite{supplementary}.

Having obtained $C_{\mathrm{mag}}(T)$, the magnetic entropy associated with the ordering of the Cr moments was then estimated to be 9.7~J~mol$^{-1}$~K$^{-1}$, a little below the expected value of $R\ln(2S+1) = 11.53$~J~mol$^{-1}$~K$^{-1}$ for a Cr$^{3+}$ ion with $S=\frac{3}{2}$ (see Supplemental Material~\cite{supplementary}). This analysis reveals that a large fraction of the magnetic entropy associated with the magnetic ordering is released well below $T_{\mathrm{N}}$, leading to the small anomaly in $C_{\mathrm{p}}(T)$ at the transition. Notably, both the absolute value of the heat capacity and the magnitude of the anomaly at the transition, closely match those reported for the related compound Cr$_{1/3}$TaS$_2$~\cite{Xie2025}, suggesting a similar strength of magnetic exchange interactions and comparable phonon spectra between the two systems, consistent with their closely related crystal and possibly magnetic structures.

	\subsection{Magnetization}
        \label{Mag}
        
  \begin{figure}
		\centering
		\includegraphics[width= 1.0\columnwidth]{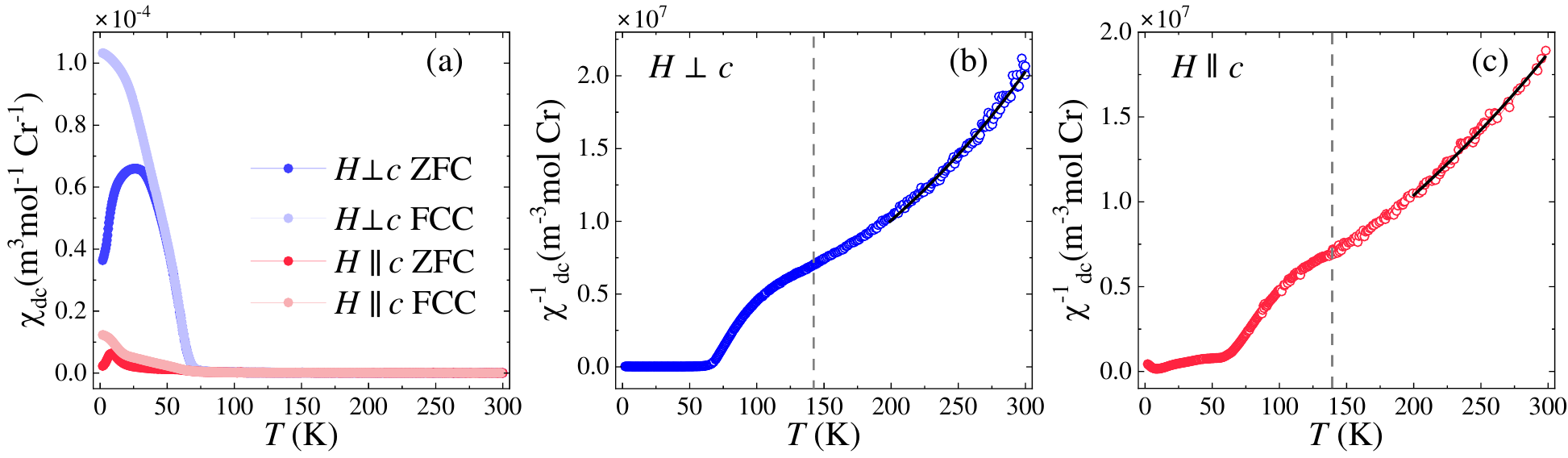}
		\caption {(a) Zero-field-cooled (ZFC) and field-cooled (FC) $\chi(T)$ measured under an applied field of $\mu_0H = 10$~mT for $H \perp c$ and $H \parallel c$. Inverse magnetic susceptibility, $\chi^{-1}$, as a function of temperature for a Cr$_{1/4}$NbS$_{2}$ single crystal with (b) $H \perp c$ and (c) $H \parallel c$. The open circles represent the experimental data and the black solid lines denote the Curie-Weiss fits performed in the temperature range 200--300~K. The grey dashed line denotes the region of change of slope in $\chi^{-1}(T)$ near 141 K.} 
	\label{MT}
 \end{figure}

The magnetic susceptibility as a function of temperature for a single crystal of Cr$_{1/4}$NbS$_2$ for two crystallographic orientations, $H \perp c$ and $H \parallel c$, are shown in Fig.~\ref{MT}(a). At lower temperatures the data are dominated by a ferromagnetic component with an ordering temperature, $T_{\mathrm{C}}$, of approximately 75~K. Cr$_{1/4}$NbS$_2$ is expected to exhibit an antiferromagnetic transition close to 141~K, as evident from the heat capacity (see Sec.~\ref{HC}) and electrical resistivity (see Sec.~\ref{Transport}) measurements. It is therefore worth pausing here to explain the origin of this ferromagnetic signal.

In intercalated Cr$_{x}$NbS$_2$ compounds, the Cr atoms occupy the van der Waals gap between the 2H-{NbS$_2$} layers. The magnetic ordering temperatures and the nature of the magnetic ground state are strongly dependent on the Cr occupancy $x$. For the $x = 1/3$ compound, the exchange interactions between the Cr moments stabilize long-range ferromagnetic order in the temperature range of 111--127~K~\cite{CAO2020100080}, below which a rich variety of magnetic phases have been reported, including chiral helimagnetic order, a field-induced chiral soliton lattice, and forced ferromagnetism~\cite{PhysRevMaterials.6.024407}. The $T_{\mathrm{C}}$ in these compounds is highly sensitive to disorder effects, which have been shown to suppress $T_{\mathrm{C}}$~\cite{PhysRevB.91.184205, 10.1063/5.0101351}. The ferromagnetic signal observed in the magnetic susceptibility [Fig.~\ref{MT}(a)] originates from Cr$_{1/3}$NbS$_{2}$ intergrowths. The onset of this ferromagnetic signal occurs at $\sim$75~K, which is much lower than the intrinsic $T_{\mathrm{C}}$ of bulk Cr$_{1/3}$NbS$_{2}$. This suppression of $T_{\mathrm{C}}$ is attributed to the high degree of structural disorder inherent to these intergrowths, which are spatially confined within small domains embedded in the majority Cr$_{1/4}$NbS$_{2}$ matrix. The fact that $\chi(T)$ below 75~K is an order of magnitude lower than that expected for bulk Cr$_{1/3}$NbS$_{2}$~\cite{PhysRevMaterials.6.024407} (see Supplemental Material for a direct comparison between the $\chi(T)$ for Cr$_{1/4}$NbS$_{2}$ and Cr$_{1/3}$NbS$_{2}$~\cite{supplementary}), corroborates the view that the Cr$_{1/3}$NbS$_{2}$ phase is present only as a minority intergrowth within a Cr$_{1/4}$NbS$_{2}$ matrix. The presence of intergrowths appears to be an intrinsic and unavoidable feature of these systems in general e.g.,  Cr$_{1/4}X$S$_{2}$ crystals, with similar observations reported for the related compound Cr$_{1/4}$TaS$_{2}$, where minority Cr$_{1/3}$TaS$_{2}$ domains also dominate the magnetization measurements~\cite{Xie2025}.

The $\chi^{-1}(T)$ curves reveal a clear change of slope near 141~K, as shown in Figs.~\ref{MT}(b, c), reflecting the intrinsic antiferromagnetic behavior of the Cr$_{1/4}$NbS$_{2}$ phase. Curie-Weiss fits to the modified Curie-Weiss equation, $\chi(T) = \frac{C}{T - \theta_{\mathrm{CW}}}+ \chi_{0}$ are also shown in Figs.~\ref{MT}(b, c). The parameters extracted from the fitting are given in Table~\ref{cw_fitting}. At high-temperature $\chi^{-1}(T)$ for $H$ applied parallel and perpendicular to the $c$-axis yields an effective magnetic moment $\mu_{\mathrm{eff}}$ of $\sim 3.8~\mu_{\mathrm{B}}$, consistent with the expected value for Cr$^{3+}$ in the $S = 3/2$ state ($\mu_{\mathrm{eff}}^{\mathrm{calc}} = 3.87~\mu_{\mathrm{B}}$). The Curie--Weiss temperatures obtained for $H \perp c$ and $H \parallel c$ are $\theta_{\mathrm{CW}}^{\perp} \sim +32$~K and $\theta_{\mathrm{CW}}^{\parallel} \sim +15$~K, respectively. The positive values of $\theta_{\mathrm{CW}}$ in both orientations are notable given that bulk Cr$_{1/4}$NbS$_2$ is an antiferromagnet with $T_{\mathrm{N}} = 141$~K. A similar behavior is observed in many magnetic systems, for example, in A-type antiferromagnets with in-plane ferromagnetic order and antiferromagnet coupling between planes~\cite{Zhou99, Wollan55}. More specifically, in the family of two-dimensional magnetic van der Waals materials, for example, single crystal CrCl$_3$ orders antiferromagnetically below 17~K, yet the Curie--Weiss temperatures are positive for fields applied both in- an out-of-plane due to ferromagnetic interactions that dominate in the paramagnetic state~\cite{PhysRevMaterials.1.014001, PhysRevB.102.014424}. Competing ferromagnetic and antiferromagnetic interactions yielding a positive $\theta_{\mathrm{CW}}$ have also been reported in other members of the transition-metal-intercalated transition metal dichalcogenide family, including V$_{1/4}$NbSe$_2$~\cite{han2026anomalous}, Co$_{0.26}$TaS$_2$~\cite{liu2022electrical}, and Cr$_{1/4}$TaS$_2$~\cite{Xie2025}.  We also note that the small volume of ferromagnetic Cr$_{1/3}$NbS$_2$ intergrowths, could contribute, in part, to the positive $\theta_{\mathrm{CW}}$ values.

\begin{table}[h]
\caption{Curie-Weiss parameters for Cr$_{1/4}$NbS$_{2}$ from fits made to $\chi^{-1}(T)$ in the temperature range 200--300~K for two crystallographic orientations, $H \perp c$ and $H \parallel c$.}
\label{cw_fitting}
\centering
\begin{tabular*}{\columnwidth}{@{\extracolsep{\fill}}lccc}
\hline\hline
 & $\chi_{0}$ & $\mu_{\mathrm{eff}}$ & $\theta_{\mathrm{CW}}$ \\
Orientation & (m$^{3}$ mol$_{\mathrm{Cr}}^{-1}$) & ($\mu_{\mathrm{B}}/$f.u.) & (K) \\
\hline
$H \perp c$ & $-3.51\times10^{-8}$ & $3.80(2)$ & $32(2)$ \\[4pt]
$H \parallel c$ & $-2.60\times10^{-8}$ & $3.79(2)$ & $15(2)$ \\
\hline\hline
\end{tabular*}
\end{table}

\begin{figure}
		\centering
		\includegraphics[width= 0.5\columnwidth]{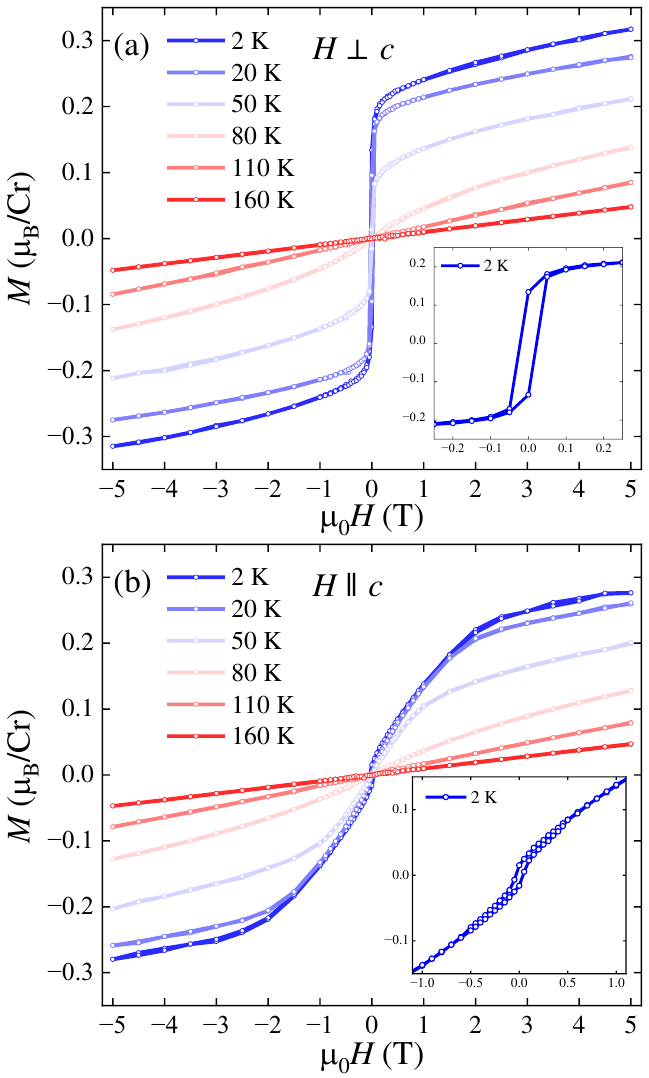}
		\caption {Magnetization as a function of applied field measured at various temperatures for (a) $H \perp c$ and (b) $H \parallel c$. The inset shows an enlarged view of the $M$ vs $H$ data at 2~K.} 
	\label{MH}
 \end{figure}

Figures~\ref{MH}(a, b) show the magnetization as a function of applied field for a single crystal of Cr$_{1/4}$NbS$_2$ for two different crystallographic orientations. For $H \perp c$, as shown in Fig.~\ref{MH}(a), the magnetization does not saturate in fields up to 5~T and instead increases linearly with field. The maximum moment achieved at 5~T is 0.31 $\mu_\mathrm{B}$/f.u., which is an order of magnitude lower than the saturation moment expected for Cr$^{3+}$ (3 $\mu_\mathrm{B}$/f.u.), consistent with the compensated antiferromagnetic ordering of the sample. The linear behavior persists down to 80~K, below which a slight $S$-shaped feature emerges with negligible hysteresis. At 2~K, the magnetization shows signs of saturation around 0.2 $\mu_\mathrm{B}$/f.u., accompanied by a hysteresis loop with a coercivity of 23~mT, before continuing to increase linearly with field. The hysteresis behavior, highlighted in the inset of Fig.~\ref{MH}(a), arises from the ferromagnetic Cr$_{1/3}$NbS$_2$ intergrowths in the sample. Assuming that the signal of 0.2 $\mu_\mathrm{B}$/f.u. arises solely from these intergrowths, an upper bound for the impurity fraction can be estimated from $\frac{M_\mathrm{sample}}{M_\mathrm{sat}}$, normalized by the respective molecular weights, of approximately 5--6\%. This is strictly an upper bound, as the magnetization in fact has several contributions including the antiferromagnetic response from the main Cr$_{1/4}$NbS$_2$ phase which increases linearly with the applied field and possibly an intrinsic weak ferromagnetic component, analogous to that reported in $\alpha$ - MnTe~\cite{WU2025173106}.

A similar behavior is observed for $H \parallel c$. The $M$~vs~$H$ response is linear above 80~K, and below 80~K the curve evolves into an $S$-shaped profile, eventually developing a clear hysteresis at 2~K, as shown in the inset of Fig.~\ref{MH}(b). The anisotropy observed in the hysteresis loops between the $H \perp c$ and $H \parallel c$ orientations is likely a consequence of the intrinsic magnetic anisotropy of the Cr$_{1/3}$NbS$_2$ intergrowths. An analogous behavior has been reported for Cr$_{1/4}$TaS$_2$~\cite{Xie2025}.

 \subsection{Magnetotransport}
	 \label{Transport}

\begin{figure}
		\centering
		\includegraphics[width= 1.0\columnwidth]{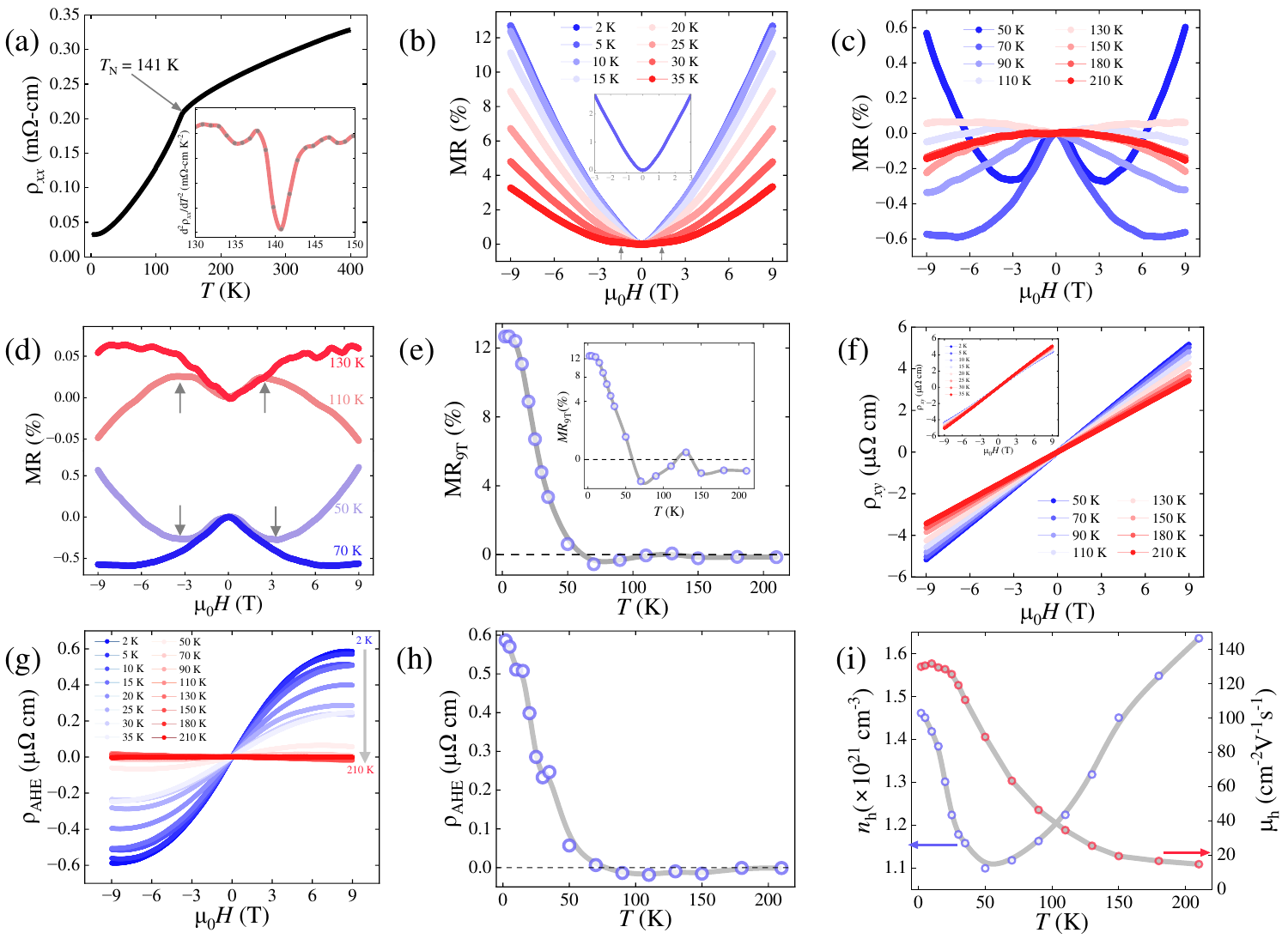}
		\caption {Magnetotransport measurements on Cr$_{1/4}$NbS$_{2}$. (a) Temperature dependence of $\rho_{xx}(T)$ measured under $H \parallel c$-axis. The inset shows the second derivative $\mathrm{d}^{2}\rho_{xx}(T)/\mathrm{d}T^{2}$ as a function of temperature, displaying a sharp dip at the N\'{e}el temperature $T_{\mathrm{N}} = 141$~K. (b) Magnetoresistance (MR) as a function of $\mu_{0}H$ for temperatures ranging from 2 to 35~K. the gray arrow indicates the kink around 3~T. The inset highlights the linear field dependence of MR below 3~T at $2$~K. (c) MR as a function of $\mu_{0}H$ for temperatures in the range 50 to 210~K. (d) MR as a function of $\mu_{0}H$ at selected temperatures of 50, 70, 110, and 130~K, with grey arrows marking the anomalous features. (e) Temperature dependence of MR evaluated at $\mu_{0}H = 9$~T. The dashed line indicates the $\mathrm{MR}=0$. The inset shows the same data on a square root scale to highlight the behavior close to $T_{\mathrm{N}}$. (f) Hall resistivity $\rho_{xy}$ as a function of $\mu_{0}H$ for temperatures in the range 50 to 210~K. The inset shows the corresponding low-temperature data for 2 to 35~K. (g) Anomalous Hall resistivity as a function of $\mu_{0}H$ at different temperatures. (h) Variation of AHE at 9~T as a function of temperature. (i) Carrier concentration $n_{\mathrm{H}}$ (left axis) and carrier mobility $\mu_{\mathrm{H}}$ (right axis) as a function of temperature.} 
	\label{MTR}
 \end{figure}

Magnetotransport measurements were conducted using a five-point wire configuration to simultaneously measure the longitudinal resistivity $\rho_{xx}$ and transverse Hall resistivity $\rho_{xy}$, with $H \parallel c$ and $I \perp c$. Both $\rho_{xx}$ and $\rho_{xy}$ are symmetrized and antisymmetrized, respectively, before analysis; the process is summarized in Supplementary Material~\cite{supplementary}. Figure~\ref{MTR}(a) shows the temperature dependence of the zero-field longitudinal resistivity. The curve exhibits metallic behavior with a sharp drop at $\sim 141$~K (the inset in Fig.~\ref{MTR}(a) shows a pronounced dip in $\mathrm{d}^2\rho_{xx}(T)/\mathrm{d}T^2$ versus temperature around this temperature), attributed to the suppression of spin disorder scattering as the system undergoes a paramagnetic to antiferromagnetic phase transition ~\cite{PhysRevB.2.4613}, establishing the N\'{e}el temperature of Cr$_{1/4}$NbS$_2$ at $T_{\mathrm{N}} \sim 141(1)$~K. The residual resistivity ratio, $\mathrm{RRR} = \rho(400~\mathrm{K})/\rho(2~\mathrm{K}) \sim 10$, is comparable to that reported for the related compound Cr$_{1/4}$TaS$_2$~\cite{Xie2025}. This value is reasonably high for the broader family of transition-metal-intercalated dichalcogenides, where single crystals typically exhibit RRR values below 10~\cite{c87z-wrhj, 6l9v-4gg8}, and in several cases as low as 2--3~\cite{Ray2025, PhysRevB.84.075157, Regmi2025}. This, therefore, confirms the high quality of the Cr$_{1/4}$NbS$_2$ single crystals investigated in this work. Figures~\ref{MTR}(b-e) present the magnetoresistance (MR) of the sample as a function of applied magnetic field, defined as $\mathrm{MR} = \frac{\rho(H) - \rho(0)}{\rho(0)} \times 100\%$, where $\rho(H)$ is the longitudinal resistivity at an applied magnetic field $H$ and $\rho(0)$ is the zero-field resistivity. The MR behavior evolves significantly with temperature and can be categorized into distinct regimes as described below. At low temperatures ($T \leq 30$~K), the MR exhibits a nearly parabolic field dependence ($\mathrm{MR} \propto H^2$) over the entire field range as shown in  Fig.~\ref{MTR}(b), with a slight deviation towards linear behavior at low fields below 2.5~T [see inset of  Fig.~\ref{MTR}(b)]. This quadratic field dependence is characteristic of multi-carrier transport within the semiclassical Boltzmann transport theory~\cite{pippard1989magnetoresistance}. A deviation from parabolic behavior occurs above 30~K.  At 35~K, altough the parabolic behavior of the MR persists, the curve develops two kink-like features near $\pm 3.3$~T [gray arrows in Fig.~\ref{MTR}(b)]. The higher temperature data (above 35~K) is shown in Fig.~\ref{MTR}(c), and at 50~K, these kinks evolve into two prominent positive cusps centered at $\sim \pm 3.3$~T; the MR becomes negative ($\sim -0.3\%$) at low fields, rises through the cusps, and eventually recovers to a positive value of $\sim 0.6\%$ at $9$~T. At $70$~K, the MR becomes entirely negative, following a linear field dependence and reaching $\sim -0.6\%$, before appearing to saturate above 7~T. The curve at 70~K displays an inverted curvature compared to the 50~K isotherm, with a positive cusp appearing at the same field near 3.3~T. Upon further warming to $110$~K, the MR continues to remain negative($\sim -0.05\%$) rises through the cusp with increasing field, eventually reaching a positive value of  $\sim 0.03\%$. At 130~K, the MR exhibits a feature that is the mirror image of that observed at 70~K. The field dependence of the curvature of MR at 50, 70, 110 and 130~K temperatures is magnified in Fig.~\ref{MTR}(d). Further details of the evolution of MR($H$) at different temperatures are provided in the Supplemental Material~\cite{supplementary}. Above 130~K, the MR reverts back to a parabolic behavior, however the MR inverted as compared to that at 2~K.  The temperature evolution of the magnetoresistance at 9\,T is summarized in Fig.~\ref{MTR}(e). The MR reaches a maximum value of $\sim 12.7\%$ at 2\,K, and decreases monotonically with increasing temperature up to 50\,K, beyond which it changes sign and becomes negative. The magnitude of the negative MR grows upon further warming, reaching a pronounced dip near 70\,K, before recovering toward zero. A further suppression is observed at the N\'{e}el temperature $T_{\mathrm{N}} = 141$\,K, above which the MR saturates and remains nearly temperature independent above 150\,K. The non-monotonic evolution of MR with both field and temperature, including the sign changes and cusp-like features, is indicative of quantum diffusive transport phenomena commonly observed in disordered metals~\cite{PhysRevB.99.045119}, dilute magnetic semiconductors~\cite{PhysRevLett.99.116803}, or topological insulators~\cite{PhysRevLett.108.036805} arising from constructive and destructive interference between electron trajectories scattered by disorder or spin textures leading to  weak localization and weak anti-localization effects~\cite{Altshuler1985}.

The Hall resistivity at several temperatures ranging from 2 to 210~K is shown in Fig.~\ref{MTR}(f). The total Hall resistivity $\rho_{xy}$ in magnetic materials is conventionally decomposed as $\rho_{xy} = R_0 \mu_0 H  + \rho_{xy}^{\mathrm{AHE}}$, where the first term is the ordinary Hall effect (OHE) arising from the Lorentz force on the conduction electrons~\cite{Hall1879}, and the second term is the contribution from the anomalous Hall effect (AHE) arising due to intrinsic (Berry curvature) and extrinsic (skew-scattering and side-jump) mechanisms~\cite{RevModPhys.82.1539}. In magnetic conductors with a large resistivity, the AHE is dominated by the intrinsic and side-jump contributions~\cite{Nakatsuji2015, PhysRev.95.1154}. To extract the individual Hall contributions, $R_0$ is first obtained by fitting the high-field region (8--9\,T) of $\rho_{xy}$, where the response is dominated by the linear OHE. The carrier concentration $n = 1/R_0 e$ and carrier mobility $\mu = 1/ne\rho_{xx}$ extracted at each temperature using $R_0$ are presented in Fig.~\ref{MTR}(i). The carrier concentration is $1.46 \times 10^{21}$\,cm$^{-3}$ at 2\,K, drops monotonically to $1.1 \times 10^{21}$\,cm$^{-3}$ around 50\,K, and recovers to $1.6 \times 10^{21}$\,cm$^{-3}$ at 210~K, tracing a well-defined dip centered at 50\,K. The carrier mobility obtained at 2\,K is $\mu = 129$\,cm$^{2}$V$^{-1}$s$^{-1}$. The obtained $R_0$ is then used to compute and subtract the OHE contribution $R_0\mu_0 H$ from the total Hall resistivity, yielding the anomalous Hall contributions $\rho_{xy}^{\mathrm{AHE}}$. As a consistency check, we assume $\rho_{xy}^{\mathrm{AHE}}=S_{\mathrm{H}}\,\rho_{xx}^{2}\, M$ and plot $\rho_{xy}(H)/\mu_0 H$ as a function of $\rho_{xx} (H)^{2}M(H)/\mu_0 H$, where the intercept yields $R_0$ and the slope yields the AHE scaling factor $S_{\mathrm{H}}$. The AHE estimated using $R_0$ from this linear fit is in good agreement with that obtained from the high-field fitting. The first method is then applied systematically across all measured temperatures and the resulting AHE contribution as a function of temperature is shown in Fig.~\ref{MTR}(g). The field dependence of $\rho_{xy}^{\mathrm{AHE}}$ closely mirrors the magnetization for fields applied along the same direction [see Fig.~\ref{MH}(b)]. The evolution of $\rho_{xy}^{\mathrm{AHE}}$ at 9~T as a function of temperature is shown in Fig.~\ref{MTR}(h). The $\rho_{xy}^{\mathrm{AHE}}$ is a maximum ($\sim 0.59 \,\mu\Omega\mathrm{~cm}$) at the lowest measured temperature (2~K) and monotonically reduces with increasing temperature, becomes negative above 70~K with a minimum around 125~K, before becoming zero above 150~K ($T >T_{\mathrm{N}}$). The change in sign of $\rho_{xy}^{\mathrm{AHE}}$ above 50\,K, and disappearance of the AHE above 150~K coincide with the unusual features observed in MR including the flattening of the MR response above 50\,K, the emergence of weak antilocalization at 50\,K, and the anomaly in the carrier concentration at 50\,K, strongly suggesting that a common underlying mechanism governs the transport anomalies in this temperature regime.

The temperature and field dependence of $\rho_{xy}^{\mathrm{AHE}}$ observed in Cr$_{1/4}$NbS$_2$ differ in some respects from those reported for the related compound Cr$_{1/4}$TaS$_2$~\cite{Xie2025}. While the AHE in the Ta material is also positive at low temperatures and changes sign close the ordering temperature of what may be a Cr$_{1/3}$TaS$_2$ intergrowth, the signal in the Ta compound is an order of magnitude lower in the positive sector. The reported field dependence of the AHE is also different in Cr$_{1/4}$TaS$_2$. It is unclear if the AHE and magnetization in Cr$_{1/4}$TaS$_2$ exhibit the same field dependence, as $M(H)$ data for a field applied along the requisite direction are not shown~\cite{Xie2025}. 

Although the microscopic origins of the features in the MR and AHE of Cr$_{1/4}$NbS$_2$ remain an open question, several plausible mechanisms may account for the observed behaviour. One possibility is the scattering of charge carriers from domain walls of Cr$_{1/3}$NbS$_2$ intergrowths embedded in the host matrix. This could give rise to weak antilocalization effects in MR as discussed earlier~\cite{PhysRevB.99.045119}; scattering from such defects could also produce the unusual dip in the carrier concentration as observed in systems with defects and local breaking of time-reversal symmetry, could induce a transverse voltage~\cite{Nakatsuji2015} thereby showing a finite AHE. The AHE may also be intrinsic to the Cr$_{1/3}$NbS$_2$ phase, which is known to host a giant topological and planar Hall effect~\cite{PhysRevResearch.4.013134}. Disentangling these competing scenarios calls for further studies using local experimental probes capable of investigating the sample at microscopic level, in order to fully understand the complex interplay between magnetoresistance, Hall response, composition and morphology that governs the low-temperature transport in this system.
 
 \section{Summary \& Conclusions}
           
	\label{SC}

We report the growth of high-quality single crystals of Cr$_{1/4}$NbS$_2$ for the first time, a phase known since 1970, with the motivation of providing a platform  for exploring  altermagnetism in this material experimentally. The crystals are thoroughly characterized using Laue diffraction and single-crystal X-ray diffraction to assess their quality and establish detailed crystallographic information. EDX analysis confirms the stoichiometry of the crystals as Cr$_{0.25}$NbS$_2$.  Zero-field longitudinal resistivity and heat capacity measurements unambiguously capture an antiferromagnetic transition at $T_{\mathrm{N}} \sim 141$\,K. Magnetic measurements also reveal the presence of the unavoidable minority intergrowths of the Cr$_{1/3}$NbS$_2$ phase embedded within the majority Cr$_{1/4}$NbS$_2$ matrix. Magnetotransport measurements reveal a series of complex features in the magnetoresistance and Hall response. The magnetoresistance as a function of field exhibits a crossover from weak antilocalization at 50\,K to weak localization at 130\,K. The magnetoresistance at 9\,T, as a function of temperature, displays intricate behavior: it decreases monotonically up to 50\,K, turns negative upon further warming, and becomes positive again at higher temperatures, before eventually saturating above the N\'eel temperature. The transverse resistivity likewise shows non-trivial behavior. The carrier concentration extracted from the ordinary Hall component exhibits a pronounced dip near 50\,K, and an anomalous Hall signal is observed at low temperatures which tends to saturates above 50~K. These coincident anomalies near 50\,K point toward the occurrence of quantum diffusive transport, a phenomenon commonly observed in disordered systems; here, the underlying disorder is attributed to the minority Cr$_{1/3}$NbS$_2$ intergrowths embedded within the dominant Cr$_{1/4}$NbS$_2$ matrix. The successful crystal growth of the potential altermagnetic candidate Cr$_{1/4}$NbS$_2$ thus raises several questions regarding the origin of the unusual features observed in its magnetotransport response and warrants further investigation.

\section*{Acknowledgments}

We acknowledge Ewan Jewkes and Patrick Ruddy for their technical support. This work was financially supported by Engineering and Physical Sciences Research Council grant EP/Z535874/1: "Single Crystal Growth at Warwick" and UK Research and Innovation grant UKRI648: "Advanced materials for next-generation spintronics: The deterministic control of altermagnets". Equipment used in this work was provided by the University of Warwick’s X-ray Diffraction and Electron Microscopy Research Technology Platforms.
\newpage\clearpage
         
	\bibliography{bib}

\end{document}